\documentclass[journal]{IEEEtran}

\ifCLASSINFOpdf
\else
\fi
\usepackage{algorithm,multirow}
\usepackage{algorithmic}
\usepackage{graphicx}
\usepackage{comment}
\usepackage{cite}
\usepackage{multirow}
\usepackage{amssymb,amsthm,mathrsfs}
\usepackage{amsfonts,epsf,graphicx,times,amsmath,multirow,url,cite}
\usepackage{float}
\usepackage{flexisym}
\usepackage[capitalise]{cleveref}

\usepackage{mathtools}
\DeclarePairedDelimiter{\ceil}{\lceil}{\rceil}

\usepackage{amsthm}
\usepackage{subcaption}
\usepackage{color}

\graphicspath{{figures/}}
 \usepackage{url}
\newcommand{\sol}{\texttt{IQuCS}}

\makeatletter
\def\ps@IEEEtitlepagestyle{%
  \def\@oddfoot{\mycopyrightnotice}%
  \def\@evenfoot{}%
}
\def\mycopyrightnotice{%
  {\footnotesize978-1-6654-4331-9/21/\$31.00 ©2022 IEEE
\hfill}% <--- Change here
  \gdef\mycopyrightnotice{}% just in case
}

\begin{document}
\title{Iterative Qubits Management for Quantum  Index Searching in a Hybrid System}

\author{
\IEEEauthorblockN{
Wenrui Mu\IEEEauthorrefmark{1},
Ying Mao\IEEEauthorrefmark{1},
Long Cheng\IEEEauthorrefmark{2},
Qingle Wang\IEEEauthorrefmark{2},
Weiwen Jiang\IEEEauthorrefmark{3},
and Pin-Yu Chen\IEEEauthorrefmark{4} \\
}
\IEEEauthorblockA{\IEEEauthorrefmark{1}
Fordham University, Email: \{wmu2, ymao41\}@fordham.edu}\\
\IEEEauthorblockA{\IEEEauthorrefmark{2} North China Electric Power University
\IEEEauthorrefmark{2} Email: \{lcheng, qingle.wang\}ncepu.edu.cn}\\
\IEEEauthorblockA{\IEEEauthorrefmark{3} George Mason University 
\IEEEauthorrefmark{3} Email: wjiang8@gmu.edu}\\
\IEEEauthorblockA{\IEEEauthorrefmark{4} IBM Thomas J. Watson Research Center,
\IEEEauthorrefmark{4} Email: pin-yu.chen@ibm.com }
}

\maketitle

\begin{abstract}
Recent advances in quantum computing systems attract tremendous attention. Commercial companies, such as IBM, Amazon, and IonQ, have started to provide access to noisy intermediate-scale quantum computers. Researchers and entrepreneurs attempt to deploy their applications that aim to achieve a quantum speedup. Grover's algorithm and quantum phase estimation are the foundations of many applications with the potential for such a speedup. While these algorithms, in theory, obtain marvelous performance, deploying them on existing quantum devices is a challenging task. For example, quantum phase estimation requires extra qubits and a large number of controlled operations, which are impractical due to low-qubit and noisy hardware. To fully utilize the limited onboard qubits, we propose IQuCS, which aims at index searching and counting in a quantum-classical hybrid system. IQuCS is based on Grover's algorithm. From the problem size perspective, it analyzes results and tries to filter out unlikely data points iteratively. A reduced data set is fed to the quantum computer in the next iteration. With a reduction in the problem size, IQuCS requires fewer qubits iteratively, which provides the potential for a shared computing environment. We implement IQuCS with Qiskit and conduct intensive experiments. The results demonstrate that it reduces qubits consumption by up to 66.2\%.

\end{abstract}


\section{Introduction}

In the past decade, remarkable progress has been achieved on top of advanced computing systems with various applications. 
%Novel algorithms and growing computational power have enabled a broad spectrum of usage scenarios ranging from big data processing to commercial image recognition. 
%For example, banking and financial services utilize data from various sources for risk management and fraud detection.
%Google Ads~\cite{ads} studies customer's behaviors to develop its personalize marketing strategies. These applications and services enable us a lifestyle with significantly more intelligence. 
At the backend side, these applications are powered by big data processing frameworks and cloud-optimized systems~\cite{mao2021speculative,mao2022differentiate, mao2022elastic, liu2021deep, cheng2021network}.
%, such as Hadoop~\cite{hadoop}, Spark~\cite{spark} and Flink~\cite{flink}. 
%Apache Spark~\cite{spark}, as a representative system, is an open-source unified analytics engine for large-scale data processing. It is deployed by many internet powerhouses, e.g., Netflix~\cite{netflix} and Amazon~\cite{aws}, at massive scale, collectively processing multiple petabytes of data on clusters of over 8,000 nodes. 
%In big data processing, unstructured (e.g., log data) and semi-structured (e.g., JSON records) searches are popular workloads. 
%A typical unstructured search problem would require to check $N/2$ values in average, where $N$ is the size of the dataset. 
%In a word count application ~\cite{examples} that aims to find the target and count its occurrence, it will require enumerating every single data point. 
%in system and compare the values. 
While these modern computing systems, they still requires significant computational power and network bandwidth to process a large amount of data. 
%In the post Moore's Law era, the proximity to the physical bound of semiconductor fabrication along with the increasing size of datasets raises the discussion on the future of traditional computing systems and their limitations. 
In parallel to classical computing systems, the fast development of quantum computing has pushed traditional designs to the quantum stage, which provides a promising alternative to computationally-intensive and data-hungry applications~\cite{broughton2020tensorflow, bergholm2018pennylane}. Considering the endless potential, quantum-based computing system attracts increasing attention in both industry and academia, hoping for certain systems to offer a quantum speedup. 

%Many efforts have been made in this field from both algorithmic and systemic perspectives. 
In the domain of quantum search, Lov Grover introduced a fast algorithm~\cite{grover1996fast}, which 
%utilizes quantum states. It 
speedups the searching problem quadratically by reducing the number of steps to roughly $\sqrt{n}$. 
With Grover's algorithm, the data starts out in the uniform superpositions such that the amplitudes of all data points are the same. Then, it utilizes an oracle $O$ to the data. The $O$ is defined as a "black box" function that only reflects the amplitudes of the searching targets and remains others untouched. Next, the algorithm applies another reflection that can amplify the amplitude of the targets and deamplify others. With certain rounds of this amplitude amplification process, the search targets will have significantly higher amplitudes compared to others. As these reflections are repeated iteratively, the algorithm zeros in on the specified targets.

Due to its genericity and quadratic speedup, Grover's algorithm has drawn tremendous attention since its publication.
%~\cite{grassl2016applying, zhang2020depth, byrnes2018generalized, morales2018variational}. 
Based on it, Brassard {\em et al.}~\cite{brassard1998quantum} propose a quantum counting algorithm to count the number of targets in a given dataset.
%~\cite{brassard1998quantum, aaronson2020quantum, egger2020quantum, Wie19, suzuki2020amplitude}. For example,  
It combines Grover’s and Shor’s~\cite{shor1999polynomial} quantum algorithms to count the number of targets, which can be seen as a process of Quantum Amplitude Estimation (QAE) based on  Quantum Phase Estimation (QPE). QAE serves as a fundamental basis for many algorithms~\cite{orus2019quantum, plekhanov2022variational, ramezani2020machine}. While QAE has the potential to provide significant speedups, its key component, QPE, requires additional qubits and a large number of controlled operations that make it impractical in the current Noisy Intermediate-Scale Quantum (NISQ) era.

Optimized algorithms have been proposed to remove the dependence of QPE~\cite{suzuki2020amplitude, Wie19, aaronson2020quantum, grinko2021iterative}.
With these solutions, we can, potentially, perform quantum counting and searching more efficiently on the~\cite{rao2020quantum, willsch2020benchmarking, amico2019experimental, pelofske2021sampling} NISQ quantum computers. 
However, the problem size when they invoke Grover’s algorithm is still the same in each iteration. Therefore, from the input perspective, the number of required
qubits remains.

In this project, we propose \sol, which tackles the problem from the input size point of view. It considers a data set of {\em (index, value)} pairs and solely utilizes Grover's algorithm to find the targets. Based on each Grover's iteration results, \sol~ attempts to filter out the pairs that are not likely to be the searching targets and only send the remaining data points to the next iteration. Consequently, the input size in each iteration is different, and, as a result, the required number of qubits would be reduced. With fewer qubits iteratively, \sol~ provides the potential for multi-tenant computing environment that limited qubits can be shared by multiple tasks.
The main contributions of this paper are summarized as follows.

\begin{itemize}
    \item We design and implement a quantum search algorithm in a hybrid system for the data set of {\em (index, value)} pairs. It solely relies on Grover's algorithm. 
    \item With the reduced input size in each iteration, \sol~ is able to use fewer qubits to complete the search.  
    \item We introduce CQC, Cumulative Qubit Consumption, that is an evaluation metric to judge the iterative quantum algorithms. 
    \item We conduct both simulations with Qiskit and experiments on IBM-Q. The results demonstrate that it saves qubits consumption by up to 66.2\%. Based on the results and analysis, we present lessons learned.
\end{itemize}

%while the quantum counting algorithm provides a quadratic speedup, it requires additional counting qubits.  

%From security perspective, Grassl {\em et~al.}~\cite{grassl2016applying} attempts to utilize it to implement an exhaustive key search for the Advanced Encryption Standard (AES) and analyze the quantum resource requirements.  

%With oracle, By considering $x=(x_0,x_1,...x_n)$ as a binary input, oracle is defined as a classical function $f:\{ 0,1 \}^n \rightarrow \{ 0,1 \}^m$, which takes input with $n$-bit binary and generates output with $m$-bit binary~\cite{gilyen1711optimizing}.

%For example, Qiskit, an open-source framework, provides tools for creating and manipulating quantum bits (qubits). 

%Tensorflow Quantum is a quantum machine learning library for rapid prototyping hybrid quantum-classical models. These programming frameworks enable a noise-free environment for quantum simulations on classical machines. More recently, commercial companies like IBM-Q, IonQ, Rigetti, and Honeywell provide limited public access to real quantum computers for research and education. 

\section{Related Work}

%The proximity to the physical bound of semiconductor fabrication is pushing the race in the development of quantum computing. Many giants and startups, such as IBM, Amazon, and IonQ, have begun to provide public access. 
Fundamental quantum algorithms, such as Grover's algorithm and quantum phase estimation, attract tremendous attention that aims to observe a quantum speedup on real devices in various applications.
In theory, these algorithms provide a quadratic or even, exponential speedup~\cite{babbush2021focus}, deploying them on NISQ quantum devices is a challenging task. For example, quantum phase estimation requires extra qubits and a large number of controlled operations, which are impractical due to low-qubit and noisy quantum hardware.
Improvement has been observed in quantum deep learning ~\cite{stein2022quclassi, stein2021qugan, stein2021hybrid} and big data analytics~\cite{das2022experimental,baheri2021tqea}. However, these applications are far from commercial deployment at scale in practice. For example, QuGAN~\cite{stein2021qugan} and QuClassi~\cite{stein2022quclassi} claim to provide fabulous performance in terms of model side; however, it is obtained with only 4 dimensional on the IBM-Q platform, which is because NISQ quantum computers are low-qubits (5~7 publicly available) and noisy machines. 
Many efforts have been made to ease quantum resource requirements (e.g., qubits, channels and volumes)~\cite{zhang2020depth, anikeeva2021number, wang2020prospect}. For example,  a depth optimization method is proposed in~\cite{zhang2020depth}. It utilizes multiple-stage processing, global and local Grover's operators (diffusion), and is able to achieve 20\% depth reduction. 

Variants of quantum counting and searching algorithms have been proposed to eliminate QPE~\cite{suzuki2020amplitude, Wie19, aaronson2020quantum, grinko2021iterative}, which is a qubit-expensive operation and obstacles the practical applications on NISQ machines. MLQAE~\cite{suzuki2020amplitude} attempts to reduce the qubit requirement with multiple iterations of Grover's algorithm that combines with a maximum likelihood estimation. Wie {\em et al.}~\cite{Wie19} utilizes Hadamard tests as less expensive alternatives to QPE. A simplified quantum computing algorithm that works without QPE is proposed in~\cite{aaronson2020quantum}; however, it introduces a large overhead. A recent effort, IQAE~\cite{grinko2021iterative}, can reduce the overhead through postprocessing the quantum results iteratively and only relies on Grover's operator. 
With these optimized solutions, we can, potentially, perform quantum counting and searching more efficiently on the NISQ quantum computers~\cite{willsch2020benchmarking, pelofske2021sampling}. 
However, these optimizations either focus on specific problems or complicated to implement. Furthermore, existing approaches still consider algorithms as indivisible tasks that results in the same problem size iteratively.  

In \sol, we consider a quantum search problem of {\em (index, value)} pairs. Different from the existing literature, \sol~ focuses on reducing the input data set by filtering out nonsolutions iteratively. It is achieved through quantum phase analysis on classical computers. With a reduced input in each iteration, \sol~ is able to utilize fewer number of qubits to complete the search task. 
\section{\sol~ Design}

This section discusses the problem setting and system architecture of \sol~that includes its framework, design logic, and functionalities of key modules.

%\subsection{Index Searching Problem}
In \sol, we consider a given data set of {\em  (index, value)} pairs, where indexes and values can be encoded individually. It is a common setting in big data  analytics, like semi-structured data in markup languages and {\em (key, value)} pairs in MapReduce~\cite{mapreduce}.  
%Therefore, the data set container (index, value) pairs. Specified by the users, some of the values, $\{v_i,...v_j\} \in GS$, are defined as searching targets. 
The goal is to find the indexes of the targeted values. Therefore, both indexes and values will be involved. 

We utilize Grover's search algorithm to complete the task. However, the original algorithm only amplifies the amplitude of the targeted states. It fails to determine and output the targets and indexes directly. Additionally, since both indexes and values need to be encoded, the qubit requirement is larger than traditional value-only searches. Thus, our objective is to output the original  {\em (index, value )} pairs and at meanwhile, reduce the number of required qubits.

\begin{figure}[ht]
\centering
\includegraphics[width=0.85\linewidth]{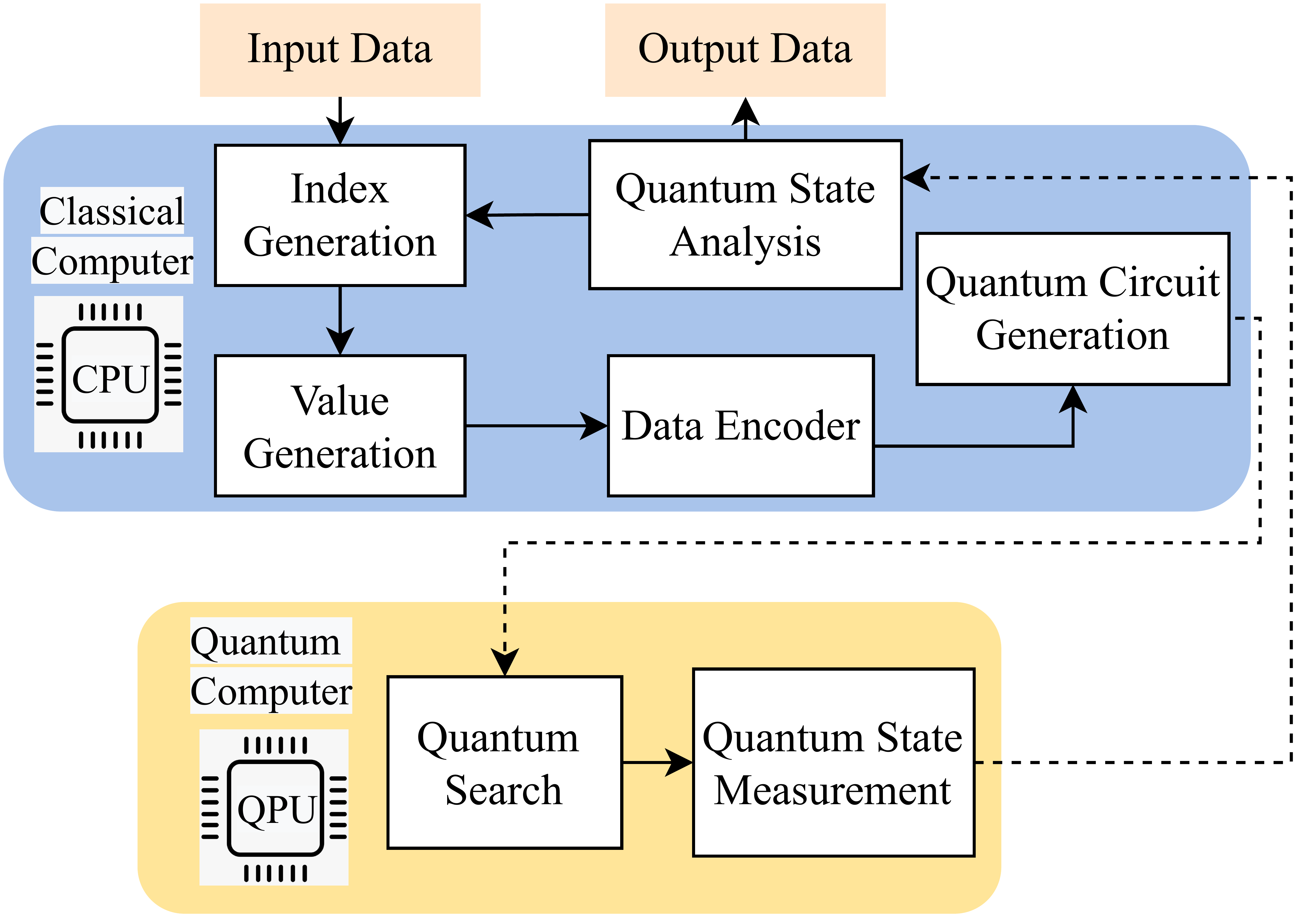}
\caption{System Architecture}
\label{fig:system}
\end{figure}

\subsection{System Design}
Figure~\ref{fig:system} illustrates the architecture of the proposed system. As a quantum-classic hybrid system, it consists of two main components, the classical computer side and the quantum computer side. The data join the system from the classical component, where Index Generation module is responsible for indexing the unstructured raw data (Algorithm \ref{alg:GenI}). In the first iteration, the generated indexes are designated as original indexes. Next, the data is sent to Value Generation module. It maps original data points to their new values in each iteration (Algorithm \ref{alg:GenV}). In our design, the values update iteratively while the search goes on. Then, the <index, value> pairs are encoded onto qubits in the Data Encoder module, and then the quantum circuit is generated for the current iteration on the classical computer. 

This circuit is passed to a quantum computer, where Grover's search algorithm is conducted with a given number of amplitude amplification. At the end of the search, the quantum states are measured and transferred back to the classical component for further processing. 

Upon receiving the results, Quantum State Analysis module is activated to perform Algorithm~\ref{alg:search}. If the algorithm finds all solutions, their original indexes will be returned. Otherwise, it conducts the filtering to ensure that only the potential solutions enter the next iteration. With this feature, the problem set is reduced when Index Generation and Value Generation modules are called in the second and following rounds. Therefore, less number of qubits are required to continue the search. Meanwhile, the mappings between current indexes and original indexes as well as current values and original values are maintained. 
\begin{table}[ht]
	\centering
	\caption{Notation Table}
	\scalebox{0.9}{
		\begin{tabular}{ | c | c |  }				
			\hline
			$I$ & Input data. \\
			\hline
			$GS$ & The set contains searching targets. \\
			\hline
			$v_i$ & The $i^{th}$ values in the data set. \\
			\hline
			$V_j$ & The input data set at iteration $j$.\\
			\hline
			$G_i$ & The input {\em (index, value)} pairs at iteration $i$. \\
		    \hline
            $NS_i$ & The set that stores nonsolutions at iteration $i$.\\
            \hline
            $PS_i$ & The set that stores potential solutions at iteration $i$. \\
            \hline
            $OI$ & The original index function that returns indexes of $V_1$. \\
            \hline
            $OV$ & The original value function that keep tracks the original values.\\
            \hline	
            $NI$ & The new index function that stores latest indexes of $v_i$.\\
            \hline
            $NV$ & The new value function that stores latest values of $v_i$.\\
            \hline
            $MpaI$ & Mappings between original indexes and current indexes.\\
            \hline
            $MpaV$ & Mappings between original values and current values.\\
            \hline
            $R_{v_i}$ & The quantum state fidelity of $v_i$.\\
            \hline
            $T_s$ & The filtering threshold. \\
            \hline
		\end{tabular}	
	}	
	\label{table:notation}
\end{table}

%%%%%%%%%%%%%%%%%%%%%%%%%%%%%%%%%%%%%%
\begin{algorithm}[!t]
\caption{Generating Indexes, GenI($V_i$)}
\begin{algorithmic}[1]
\STATE  Inputs: $V_i, j=0$
\item[]

\IF{$i=1$ for $V_i$}
    \STATE $OI({V_1}) = -1$
    \FORALL{$v_j \in V_1$}
        \STATE $OI({v_j}) = j$
        \STATE $MapI \gets [j, j]$
        \STATE Call GenV($v_i$);
        \STATE $j++$
    \ENDFOR
 %   \STATE Return $G_i$;
\item[]
 
\ELSIF{$i \neq 1$}
    \FORALL{$v_i \in V_i$}
        \IF{$OI(v_i) \neq -1$}
            \STATE $NI(v_i) = j$
            \STATE $MapI \gets [i, j]$
            \STATE Call GenV($v_i, j$);
            \STATE $j++$
        \ENDIF
    \ENDFOR
%    \STATE Return $G_i$;
\ENDIF    
\end{algorithmic}
\label{alg:GenI}
\end{algorithm}

\subsection{\sol~ Algorithms}
Searching for indexes of targeted values is a common task. Algorithm~\ref{alg:GenI} assigns indexes for the given data set iteratively and keeps mapping the original index with its current index. In the first iteration, the system calls $OI$, a function that stores Original Indexes, to set their initial values to $-1$, which indicates the not-available status (Lines 1-3). 
For every data point in $V_1$, starting from $0$, it incrementally sets indexes, stores mapping of current indexes with its original values in $MapI$ and calls Algorithm~\ref{alg:GenV} to pair each index with its corresponding data value (Lines 4-9). 

Due to the data filtering process, the input set may reduce iteratively. The system requires regenerating new indexes in every iteration and maintain the mapping between original indexes and their current values. In the $i^{th}$ iteration, 
the algorithm neglects invalid data points that indicate by negative indexes in the previous round. For valid data points, it regenerates indexes incrementally, updates the mapping of the original index, $i$, with its latest value, $j$, in $MapI$, and assigns them to $NI$ (Lines 10-14). 

Next, it invokes $GenV$ function to pair the new index with its corresponding data value (Lines 12-18). In our system, $OI$ always stores original indexes of filtered data points in $V_1$ and $NI$ is updated iteratively to store the latest indexes. With this design, the volume of indexes decreases as the search process continues. The required number of qubits is consequently reduced with a reduced number of indexes.

%the algorithm assigns a new index to data points that currently have a non-negative ($\neq 1$) index. The non-negative index 

Besides indexes, \sol~ encodes the corresponding values iteratively. With the smaller input size, it further reduce the required qubits. Our system utilized Algorithm~\ref{alg:GenV} to map original data values to their new values in each iteration. When $GenI$ calls it for the first time, $GenV$ pairs each value $v_i$ in $V_1$ with its corresponding index. The paired data is stored in $G_1$, which serves as the input for the quantum search algorithm (Lines 1-8). 

In the following iterations, it checks indexes for each data value. If $-1$ is found, this data point is marked as a nonsolution and has been filtered out in the previous round. 
$GenV$ ignores nonsolutions to the reduce input size.
The remaining data points with positive indexes are potential solutions and $GenV$ adds them set $PS_i$ (Lines 9-12). Next,
the system searches for $v_j$'s original value $v_o$ by using the function $OV$ that stores the original mapping in $v_1$ (Line 13). Then,
a rank function is employed to generate new values based on its original value, $v_o$. This rank function maps the values of potential solutions to their new values in a fixed length according to the number of elements in $PS_i$. The new values are stored in $NV$. Furthermore, $Mapv$ updates $v_o$'s latest value to $NV(v_o)$ (Line 14-16). Finally, the new index $j$ along with its corresponding new value ($NV(v_o)$) is inserted to $G_i$ that serves as the input of the next iteration (Lines 17-20).

%%%%%%%%%%%%%%%%%%%%%%%%%%%%%%%%%%%%%%
\begin{algorithm}[!t]
\caption{Generating Values (GenV)}
\begin{algorithmic}[1]
\STATE  Inputs: $V, j=0$
\item[]

\IF{$i=1$ for $V_i$}
    \FORALL{$v_j \in V_1$}
        \STATE $OV({v_i}) = v_i$
        \STATE $MapV \gets [v_i, v_i]$
        \STATE $G_1 \gets [j,~v_{i}]$
    \ENDFOR
    \STATE Return $G_i$;
\item[]
\ELSIF{$i \neq 1$}
    \FORALL{$v_j \in V_i$}
        \IF{$OI(v_j) \neq -1$}
            \STATE  $PS_i \gets v_j$
            \STATE $v_o = OV(v_j)$
            \STATE $NV(v_o) = Rank(v_o)$
            \STATE $MapV \gets [v_o,~NV(v_o)]$
            \STATE $G_i \gets [j, ~NV(v_o)]$
        \ENDIF
    \STATE Return $G_i$    
    \ENDFOR
\ENDIF    

\end{algorithmic}
\label{alg:GenV}
\end{algorithm}

%%%%%%%%%%%%%%%%%%%%%%%%%%%%%%%%%%%%%%

Based on the previous steps of $GenI$ and $GenV$, Algorithm~\ref{alg:search} performs an iterative quantum search. Initially, the input data set is sent to $GenI$, which calls $GenV$ to generate paired input set $G_1$ and maintain the original mappings. It only happens in the first iteration, when $i=1$ (Lines 1-3). 

Next, the system invokes Grover's search with $G_i$ as the input and $GS$ as the searching targeted set. When $i$ is an odd number, iteration is set to 1; otherwise, it is set to 2. This means that Grover's operator will be invoked either 1 or 2 times depending on the value $i$. (Line 4-5).
Please note that $G_i$ consists of both values and indexes in their quantum states. The resulting quantum state fidelities are stored in $R$ (Lines 6). 
By analyzing results, there are two scenarios.
\begin{itemize}
    \item We first define the mean value to be the average of all possible data points in $G_i$, which is determined by the number of encoding qubits ${\ceil{\log_2|G_i|}}$.
    When the state fidelity of $v_j$ is lower than the mean value multiplied by a threshold, $T_s$, of all possible data points in $G_i$, which deternines by the number of encoding qubits ${\ceil{\log_2|G_i|}}$, it indicates that $v_j$ is unlikely to be the solution. Therefore, the algorithm adds it to the nonsolution set $NS$ and resets its index value to $-1$. A negative index value suggests that this data point has been filtered out and will not get involved in the further iterations (Lines 7-10). 
    \item When the state fidelity of $v_j$ is higher than the mean value, the corresponding data point is a potential solution. In this case, it will be added to the $PS_i$ set for further processing (Lines 11-14).
\end{itemize}

Next, potential solutions of the current iteration are compared with the solutions of the previous iteration. There are two cases of comparison. 

\begin{itemize}
    \item If they are identical, the algorithm has converged. Then, the current indexes of all $v_i \in PS_i$ are inserted to $S$. With $S$, the algorithm finds out original indexes of all solutions (Lines 15-19).
    \item When they have a difference, it suggests that the results are not stable. The algorithm will eliminate nonsolutions from the input. Then, it invokes $GenI$ for the next iteration on a reduced dataset (Lines 20-25).   
\end{itemize}

%%%%%%%%%%%%%%%%%%%%%%%%%%%%%%%%%%%%%%
\begin{algorithm}[!t]
\caption{Iterative Quantum Search}
\begin{algorithmic}[1]
\STATE  Initialization: $I =$ Initial Inputs

\STATE $G_1 = $ GenI($I$)
\STATE i = 1
\item[]

\WHILE{true}
    \STATE GroversSearch($G_i$, $GS$, $(i+1) \mod 2 + 1$)
    \STATE Update quantum state fidelities in $R$
    \FOR {$v_j \in \{G_i\}$}
        \IF {$R_{v_{j}} < \frac{1}{2^{\ceil{\log_2|G_i|}}} \times T_s$}
            \STATE $NS_i \gets v{_j}$
            \STATE $OI(v_j) = -1$
        \ELSE
            \STATE $PS_i \gets v{_j}$
        \ENDIF    
    \ENDFOR
    
    \IF {$PS_i = PS_{i-1}$}
        \FORALL{$v_i \in PS_i$}
            \STATE $S \gets G_i(v_i)$
        \ENDFOR    
        \STATE Return $MapI(S)$
    \ELSE
        \STATE $V_i = V_i - NS_i$
        \STATE $G_i = $ GenI($V_i$)
        \STATE $i = i+1$

    \ENDIF
\ENDWHILE
\end{algorithmic}
\label{alg:search}
\end{algorithm}

\section{Evaluation}

This section presents our \sol~ implementation details and results from intensive Qiskit simulations and experiments on IBM-Q.

\subsection{Experimental Framework and Evaluation Metrics}

We implement \sol~ with Python 3.8 and IBM Qiskit Quantum Computing simulator package~\cite{qiskit}. The Aer simulator is in used as the backend to simulate a noise-free environment. The Grover's search module is constructed from Qiskit's amplitude amplifiers APIs. We set the number of shots to 12,000 and set threshold, $T_s = 0.85$.

The most common words in English~\cite{common} are encoded with their ranks into binaries, which act as values in our evaluation. For each data point, its initial index is the same as its value. Therefore, our workload is a data set of (key, value) pairs. 

We consider two types of search scenarios, (1) The values in the data set is unique; (2) There are duplicates in the data set. 
%Both single- and multiple-target settings are involved. 

The results are compared with the original Grover's search algorithm. To presume the best performance, we use {\em optimal\_num\_iterations} method~\cite{bestiteration} to calculate the number of Grover's operator invocations, which requires knowing the number of targets beforehand. Please note that this information is NOT available to \sol. To determine the targets, it utilizes the same filter as \sol. For a specific value, if its probability is higher than the mean value (Line 8 in Algorithm 3 when $i=1$) multiplies $T_s$, we assume it is a target. In the rest of this section, we use {\bf GSearch} to represent this solution.

To analyze the results, we consider two metrics: (1) Accuracy; (2) Number of invocations of Grover's operator, which is called repeatedly for amplitude amplification in Grover's algorithm; (3) Cumulative Qubit Consumption  (CQC); 

The CQC for original Grover's algorithm is straightforward since it only has one round of Grover's operator invocations. $CQC = N_q \times I$, where $N_q$ is the number of qubits to execute the algorithm, and $I$ is the calculated optimal number of Grover's operator invocations.
The VCR is defined by the equation, 
$CQC = \sum_{i=1}^{i=n} C_i \times N_{q_i}$, where $i$ is the iteration number, $C_i$ is the number of Grover's operator invocations at iteration $i$ and $N_{q_i}$ is the number of qubits at iteration $i$.

%and it is evaluated with the following series of experiments: unique dataset with one marked state, dataset with multiple unique marked states, and dataset with multiple duplicated marked states. We evaluate the algorithm on simulators and compare out results with original Grover's algorithm. In the rest of this chapter, we compare the number of iterations and the number of marked states found in different groups of experiments in detail.

\begin{figure}
\begin{subfigure}{0.23\textwidth}
  \centering
  \includegraphics[width=0.95\linewidth]{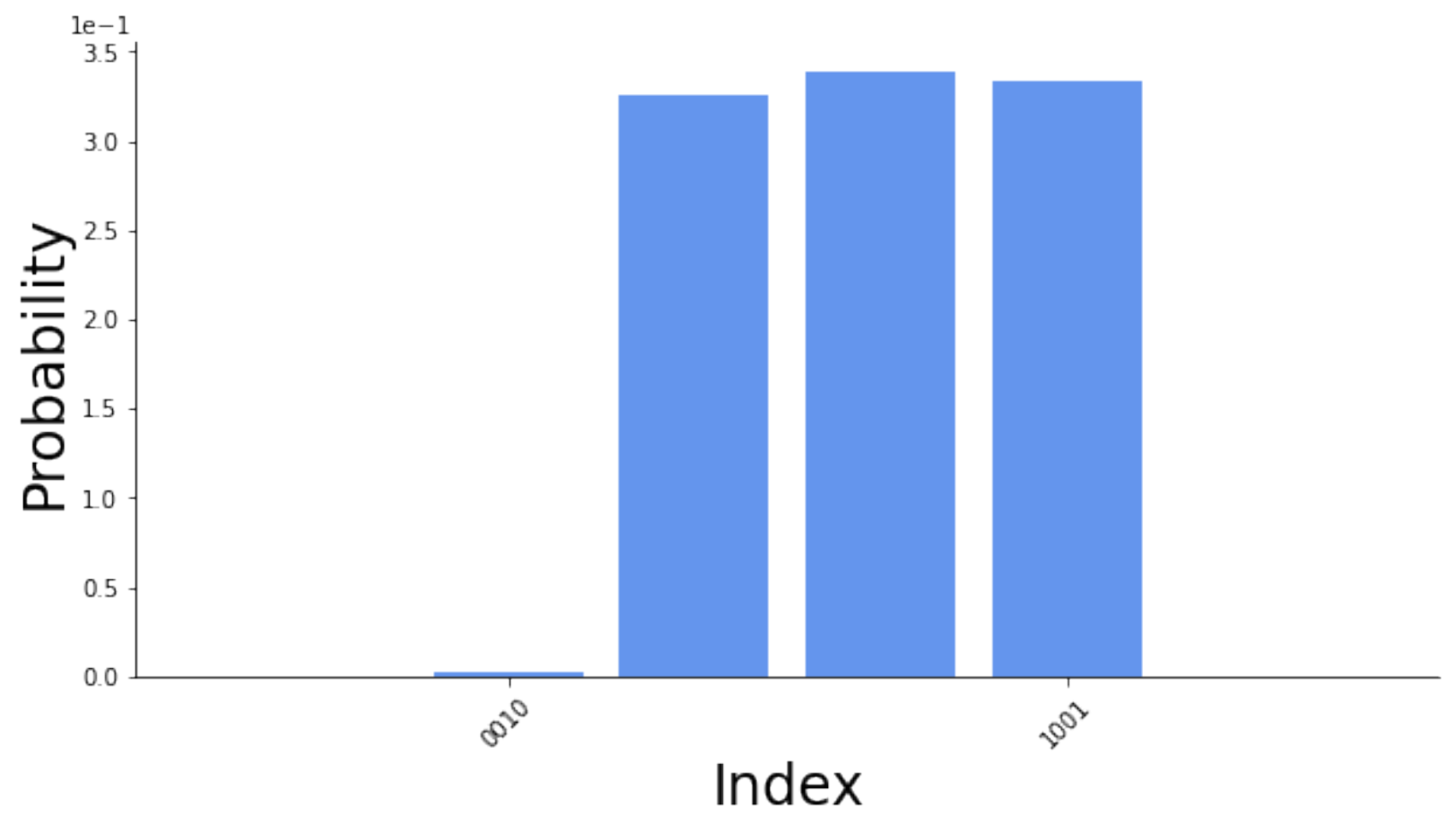} 
  \caption{GSearch: Iteration 1 - Invocation 7}
  \label{fig:grover-10-3}
\end{subfigure}
\begin{subfigure}{0.23\textwidth}
  \centering
  % include third image
  \includegraphics[width=0.95\linewidth]{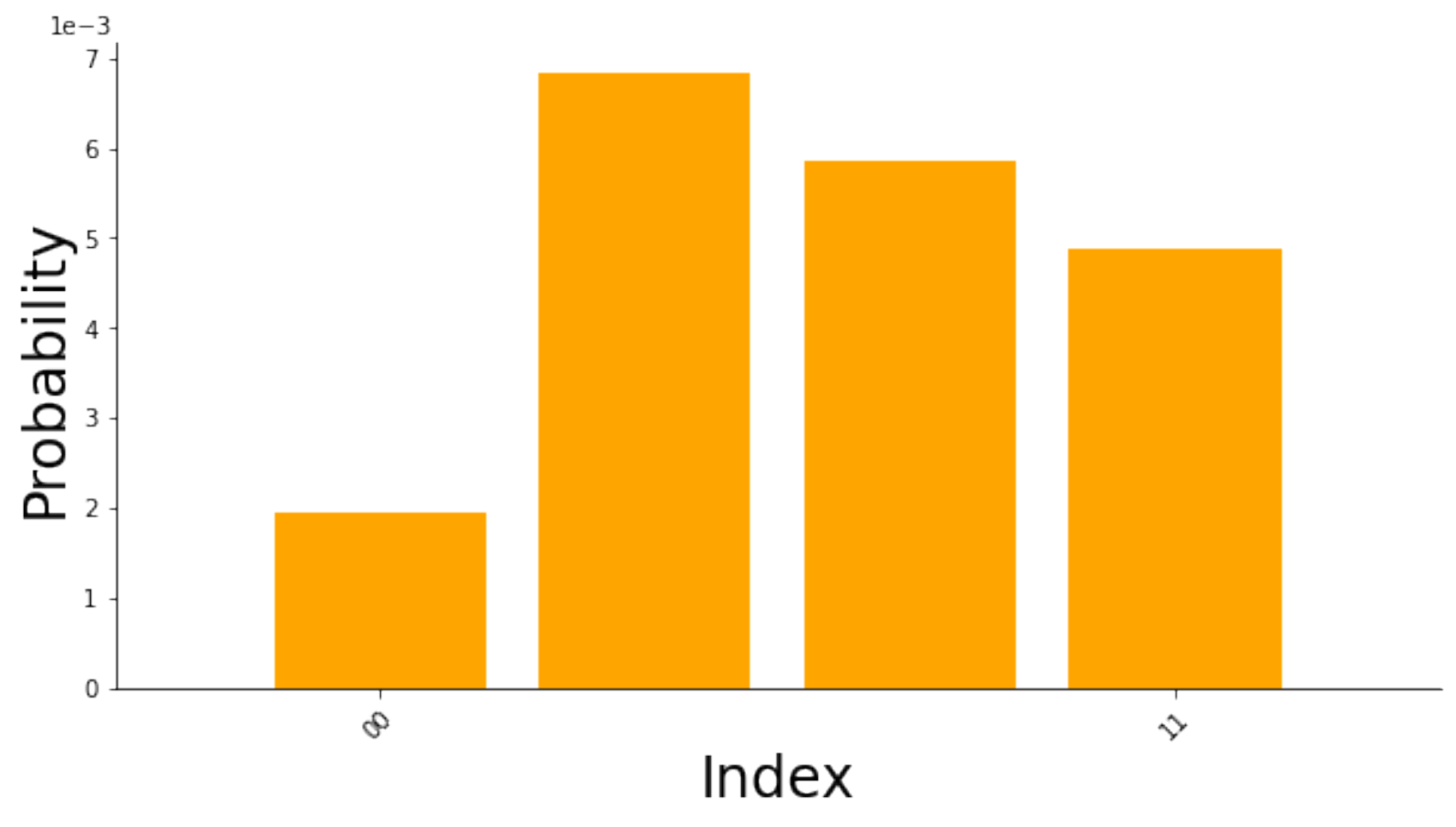}  
  \caption{\sol: Iteration 3 - Invocation 4}
  \label{fig:iqucs-10-3}
\end{subfigure}
\caption{Data set 10 - Target 3.}
\end{figure}

%%%%%%%%%%%%%%%%%%%%%%%%%%%%%%
\begin{figure}
%\begin{subfigure}{0.24\textwidth}
  \centering
  \includegraphics[width=0.75\linewidth]{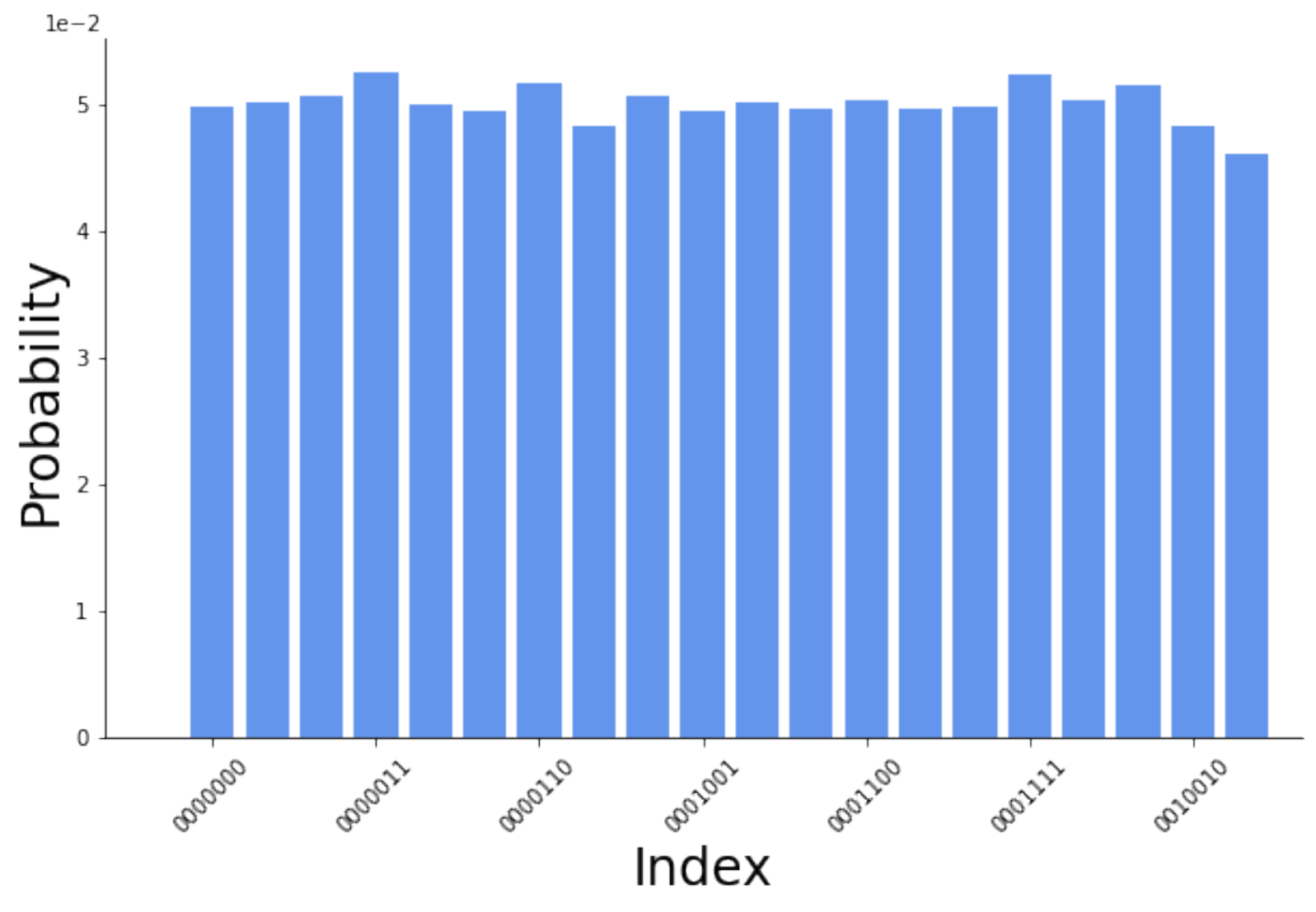} 
  \caption{Iteration 1 - Invocation 22 (GSearch)}
  \label{fig:grover-100-20-9}
%\end{subfigure}
\end{figure}

\begin{figure}
%\begin{subfigure}{0.24\textwidth}
  \centering
  \includegraphics[width=0.75\linewidth]{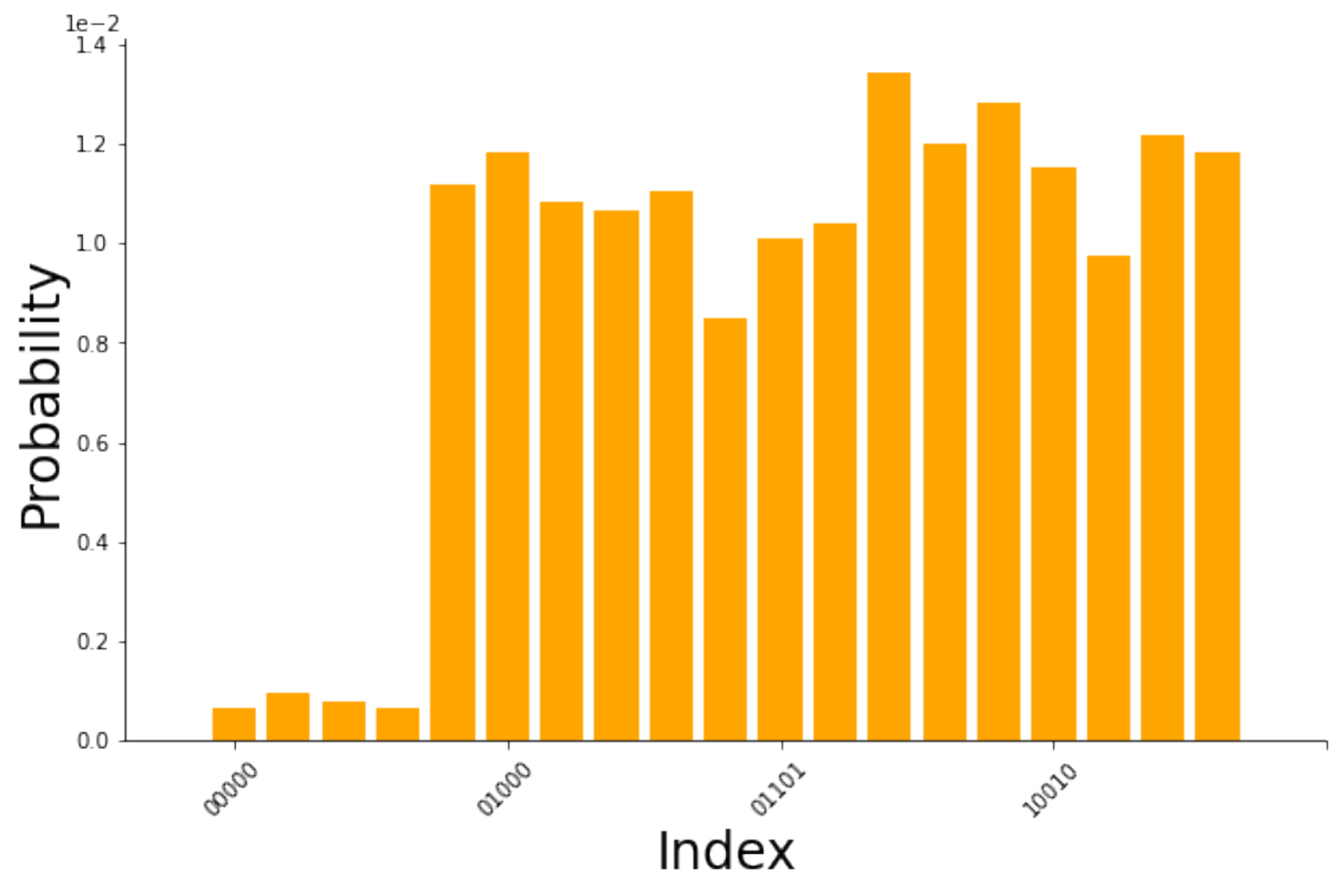}  
  \caption{Iteration 6 - Invocation 9 (\sol)}
  \label{fig:iqucs-100-20-6-final}
%\end{subfigure}
\end{figure}

%%%%%%%%%%%%%%100-20%%%%%%%%%%%%%%%%%%%%%%

\begin{figure*}
\begin{subfigure}{0.45\textwidth}
  \centering
  \includegraphics[width=0.90\linewidth]{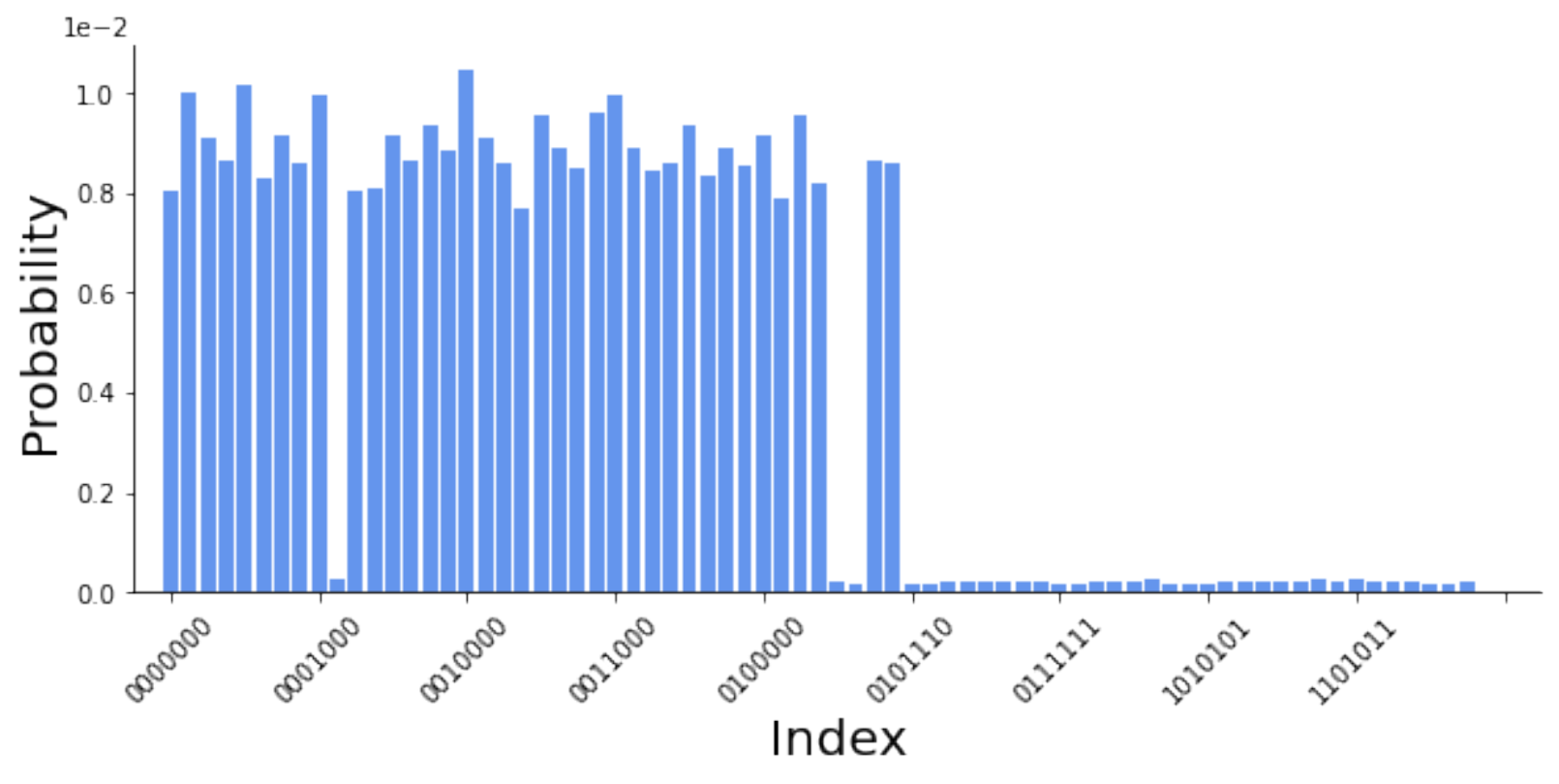}  
  \caption{Iteration 1 - Invocation 6}
  \label{fig:grover-100-40-6}
\end{subfigure}
\begin{subfigure}{0.45\textwidth}
  \centering
  % include third image
  \includegraphics[width=0.90\linewidth]{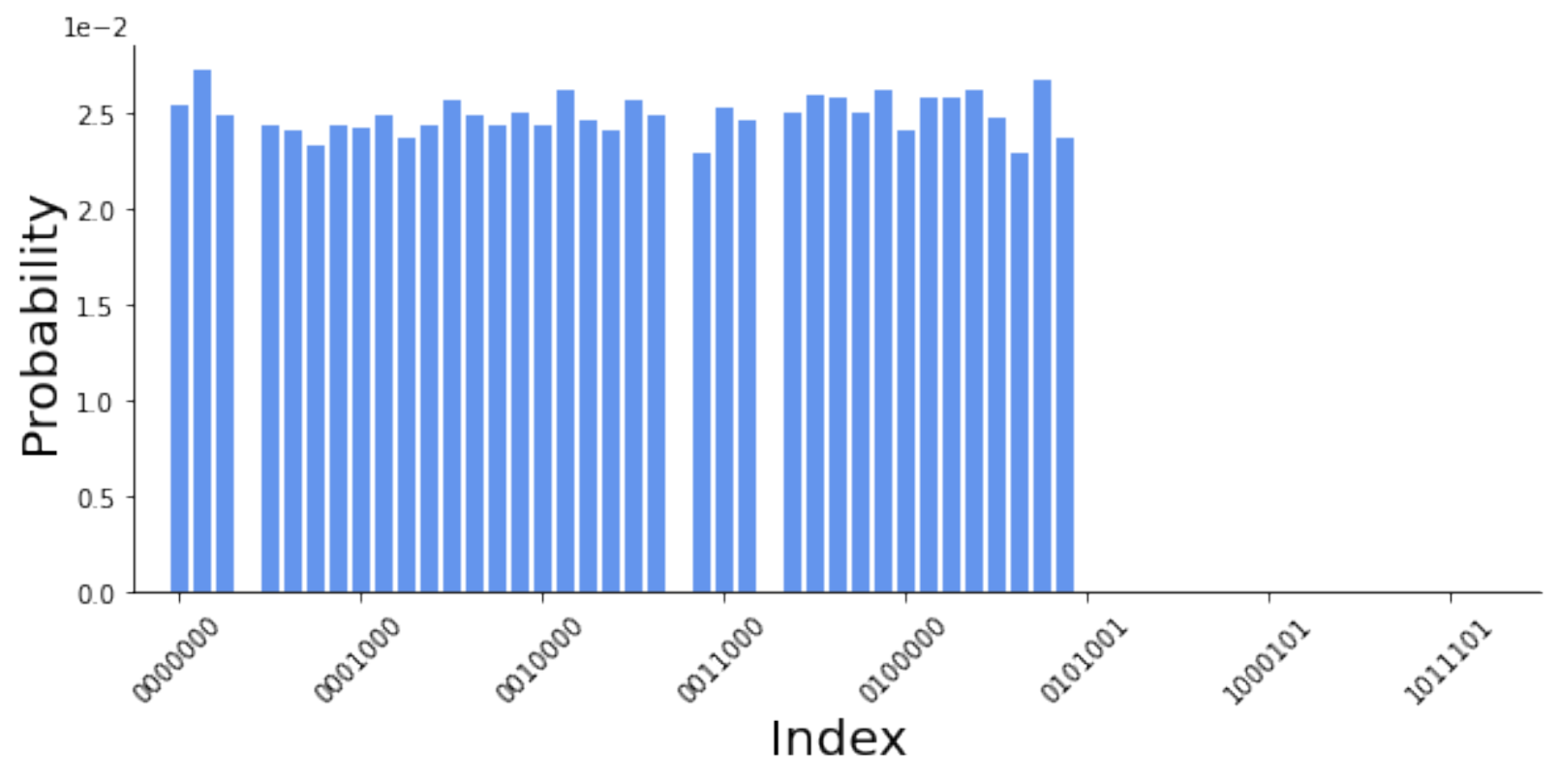}
  \caption{Iteration 1 --- Invocation 15}
  \label{fig:grover-100-40-final}
\end{subfigure}
\caption{GSearch: Data set 100 - Target 40.}
\label{fig:grover-40}
\end{figure*}

\begin{figure*}
\begin{subfigure}{0.45\textwidth}
  \centering
  \includegraphics[width=0.90\linewidth]{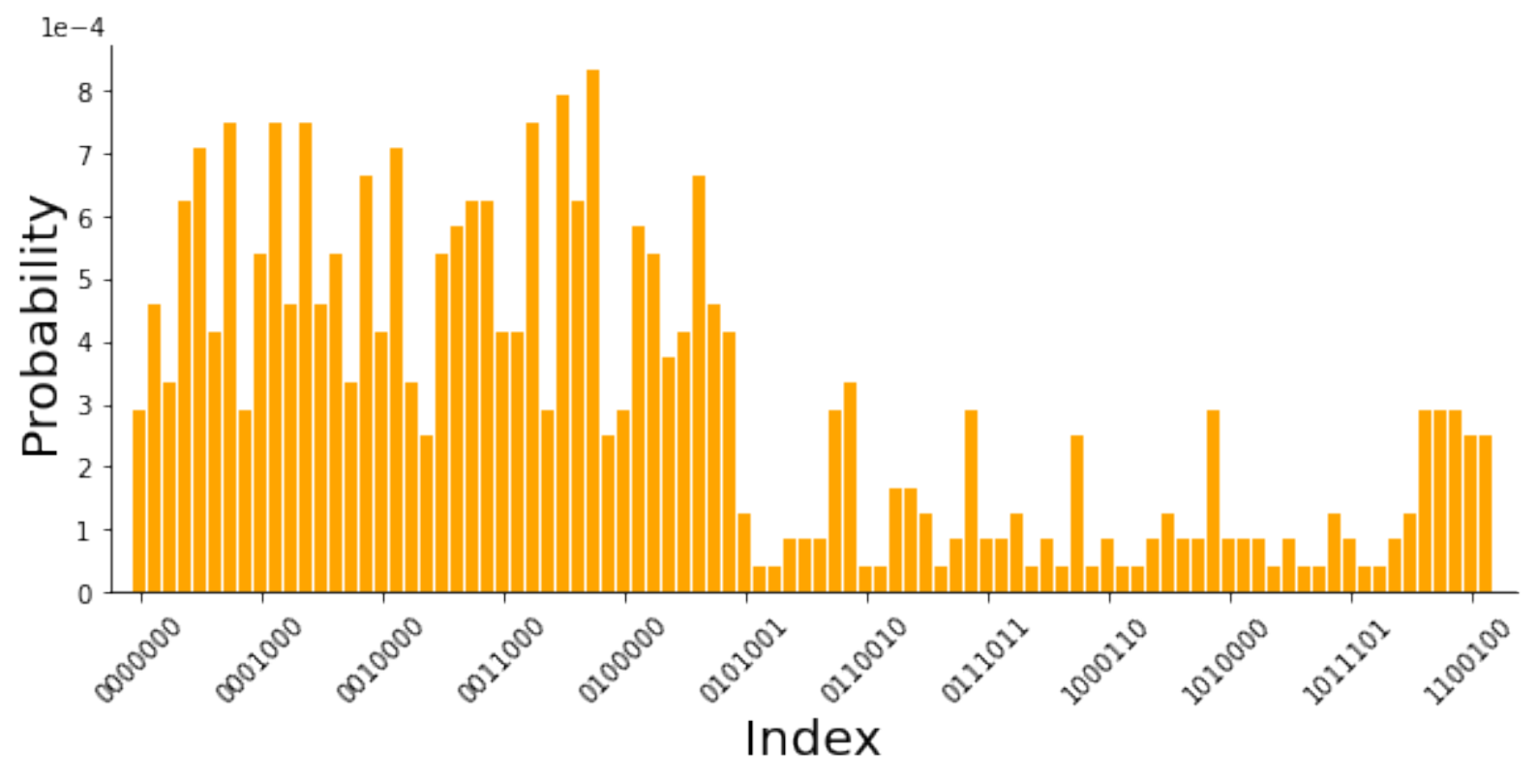}  
  \caption{Iteration 1 - Invocation 1}
  \label{fig:iqucs-100-40-1}
\end{subfigure}
\begin{subfigure}{0.45\textwidth}
  \centering
  \includegraphics[width=0.90\linewidth]{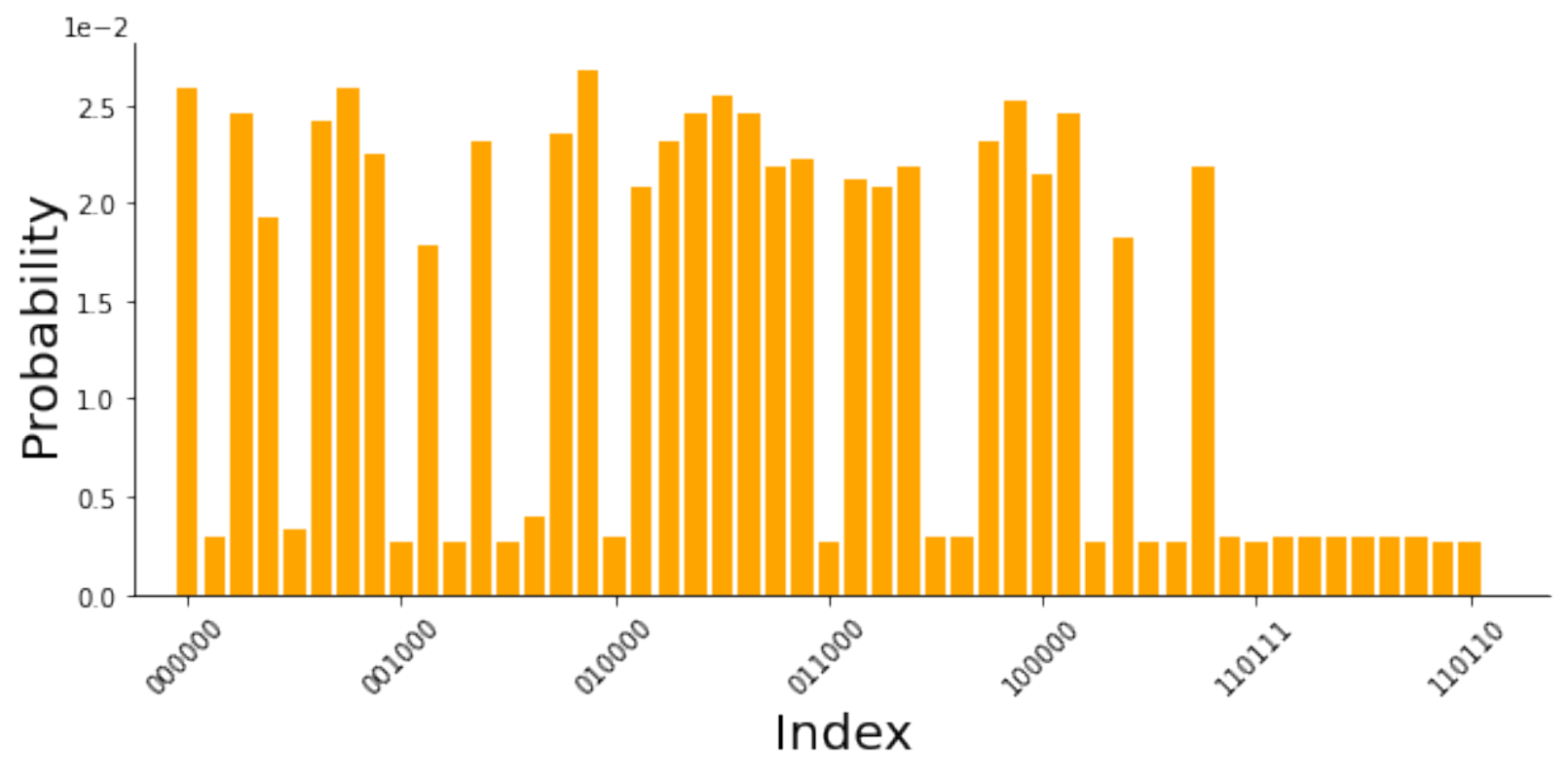}  
  \caption{Iteration 4 - Invocation 6}
  \label{fig:iqucs-100-40-4-final}
\end{subfigure}
\caption{\sol: Data set 100 - Target 40.}
\label{fig:iqucs-40}
\end{figure*}

%%%%%%%%%%%%%%%%%%%100-40%%%%%%%%%%%%%%%%%%%%%%%%%
\subsection{Data set --- 10}
In these experiments, the size of our data set is 10, which means the top 10 words are in use. We set 3 of them as our search targets.

Figure~\ref{fig:grover-10-3} and Figure~\ref{fig:iqucs-10-3} present the results by excluding data points with zero probabilities. For GSearch, its optimal invocation number is 7. Therefore, it needs to call Grover's operator 7 times. At the end, both GSearch and \sol~ can find all 3 targets. As we can see in Figure~\ref{fig:grover-10-3}, the targets' probabilities are significantly higher than others, more than 300x times. In Figure~\ref{fig:iqucs-10-3}, the difference is much smaller with \sol, where the highest probability of nonsolutions is 2.5x times lower than the lowest one among the targets. Taking a detailed look at \sol, it filters out 6 out of 10 nonsolutions, and the remaining 4 data points are sent to iteration 2. Since the problem set is reduced, the number of required qubits is also reduced, from 4 to 2 for the values. 
\sol~ successfully discovers the difference in iteration 2 and confirms them in iteration 3. When the search completes, \sol~ invoked Grover's operator $1+2+1 = 4$ times.

Both \sol~ and GSearch obtain 100\% accuracy; however, the resource consumption varies. Since the problem set of GSearch remains the same for each invocation, there is no qubits release until it finishes. The CQC for GSearch is $8\times7 = 56$. For \sol, the data points are filtered out iteratively, and thus, the number of required qubits reduces iteratively. The CQC value for \sol~is $8\times1+4\times2+4\times1 = 20$, a 64.3\% reduction.

%%%%%%%%%%%%%%%%%%%%%%%%%%%%%%%%%%%%%%%%%%%%%%%%%%%%%%%%%%%%%%%%%%%%%%%%%%%%%
\subsection{Data set --- 100}

In these experiments, our data set contains the top 100 words. We set the number of targets to 40 and 20 aiming to evaluate \sol~ in different scenarios. Please note that the number of solutions is only used to calculate the optimal invocation number of GSerch and verify the correctness for both GSearch and \sol.

{\bf 20-Targets:} Figure~\ref{fig:grover-100-20-9} and Figure~\ref{fig:iqucs-100-20-6-final} illustrate the results the 20-target experiment with Data set 100. In this case, GSearch's optimal invocation number is 22. When the search finished, GSearch discovered all 20 targets. Unfortunately, \sol~ missed 3 of them. 
In total, \sol~ executes 6 iterations to complete the search.
In the first iteration, 65 data points survive from the filter, and the other 35 values are excluded. At this moment, the remained data contains all 20 solutions. 
The same situation happens in iterations 2-4. An additional 43 of the remaining data points are filtered out. The filtered values are nonsolutions, and \sol~ made the correct decision. However, at iteration 5, another 5 are excluded, including 3 targets. The reason lies in the fact that $T_s$ value is aggressively large, which results in more data filtered in each iteration, but meanwhile, it leads to a higher probability of true negatives. Figure~\ref{fig:grover-100-20-9} plots the results of GSearch. As we can see, the number of bars are significantly less than 100, the size of the data set. This is because the probabilities of the targets are amplified with the optimal number of Grover's operators invoked. 

In terms of accuracy, GSearch achieves 100\% and \sol~ gains 97\%. When considering the qubits consumption, GSearch's CQC is 308, and \sol~ is able to reduce it to 104, a 66.2\% reduction.  

{\bf 40 Targets}: Figure~\ref{fig:grover-40} and Figure~\ref{fig:iqucs-40} present the results of 40-targets experiment. When the search completes, as shown on Figure~\ref{fig:grover-100-40-final},\ref{fig:iqucs-100-40-4-final}, both GSearch and \sol~ successfully finds all 40 targets. 
While GSerach only requires 1 iteration, its optimal invocation number is 15, which means that it calls Grover's operator 15 times. Compared with \sol, it terminates at iteration 4, which performs 6 invocations, 1, 2, 1, 2 for each specific iteration, respectively. \sol~ can reduce 60.0\% of the invocations.  Figure~\ref{fig:grover-100-40-6} plots the intermediate results of GSearch when it is at the $6^{th}$ invocation. It is able to allocate 41 targets, which is very well; however, not perfect. Taking a close look at \sol, after the first iteration, it filters out 15 nonsolutions. That is to say, the problem set is reduced to 85 in the second iteration, after which it is further reduced to 50. In the third iteration, \sol~ successfully finds all the targets. However, the system has no clues to decide the number of solutions. According to algorithms, the fourth iteration is performed, and the same 40 targets are returned, suggesting that they are all targets and the search stops.

In these experiments, both of them obtain 100\% accuracy. From a qubits consumption point of view, the CQC of GSearch is 14$\times$15 = 210, which means it calls Grover's operator 15 times, and every time it utilize 14 qubits, 7 for indexes and 7 for values. For \sol, it not only reduced the number of invocations but also reduced the number of required qubits in each iteration. The CQC is 14$\times$1 + 14$\times$2 + 12$\times$1 + 12$\times$2=78. Therefore, CQC gains a 62.9\% reduction.

\subsection{Cumulative Qubit Consumption}

\begin{figure}
  \centering
  \includegraphics[width=0.8\linewidth]{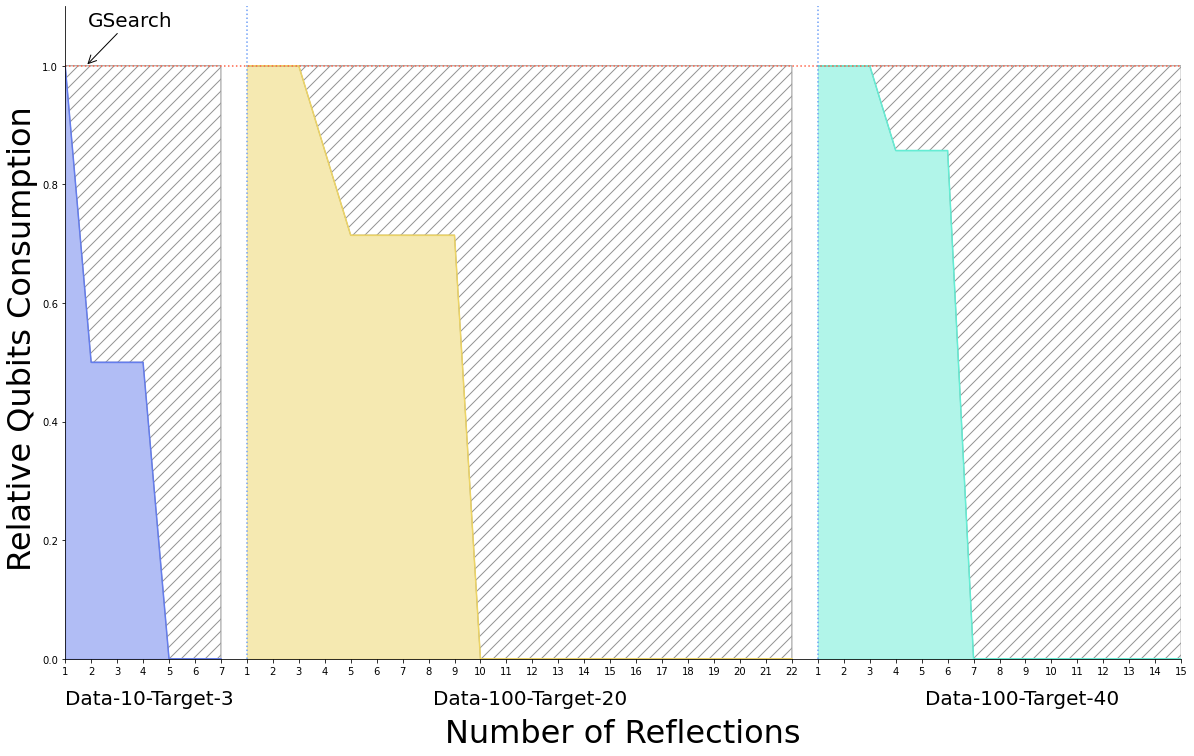} 
  \caption{Qubits Consumption Comparison}
  \label{fig:consumption}
\end{figure}

Figure~\ref{fig:consumption} presents the \sol~ qubits consumption, in relative to GSearch, of each invocation. We assume the consumption of GSearch is 1. At the first Grover's operator invocation, the consumption is always the same for both \sol~ and GSearch since they have the same initial input size. As the algorithms proceed, GSearch's input set remains, and only the probability of the individual data point updates after each call. With \sol, the input set reduces since it filters out data points iteratively. Therefore, it may require fewer qubits in the next iteration. The shadowed spaces on the figure show the saved qubit resources of \sol. The saved qubit consumption in different iterations potentially enables a multi-tenant environment such that the quantum computer can be shared by other users in second or later iterations.

\subsection{IBM-Q Experiments}

We conduct the experiments on IBM-Q quantum computers with 5-qubits, Belem, Lima and Quito, and 7-qubits Jakarta. The value-only data is considered due to limited qubits. In these experiments, we focus on the execution time and set the number of invocations to 1, 3, and 5, the number of targets to 1. Table~\ref{table2} present the results in seconds. When invocation is 1, they perform similarly since both of them has only 1 iteration. As the searching goes on with more invocations, GSearch's time cost is stable and \sol~ grows. The reason lies in the fact that \sol~ requires multiple queries to the quantum computer, which has to compile and initialize the circuits for each query that generates significant overhead. While the total number of invocations reduces, the saved time cost fails to overcome the loss of multiple initialization phases.  

\begin{comment}
\begin{table}
\centering
\begin{tabular}{ | c|c| c |}
\hline
 Machines & GSearch & \sol \\ 
 \hline
 Belem & 4.38 / 4.43 / 4.92 / 4.97 & 4.31 / 9.11 / 13.51 / 13.21 \\ 
 \hline
 Lima  & 6.28 / 6.73 / 7.15 / 7.14& 6.36 / 12.62 / 19.21 / 19.26\\
 \hline
 Quito & 4.21 / 4.56 / 4.75 / 4.82 & 4.47 / 8.42 / 13.04 / 13.12 \\
 \hline
 Jakarta & 5.85 / 6.14 / 6.16 / 6.13 & 6.01 / 11.72 / 16.98 / 16.23 \\
 \hline
\end{tabular}
 \caption{Experiments on IBM-Q.}
 \label{table2}
\end{table}
\end{comment}

\begin{table}
\centering
\begin{tabular}{ | c|c| c |}
\hline
 Machines & GSearch & \sol \\ 
 \hline
 Belem & 4.38 / 4.43 / 4.92  & 4.31 / 9.11 / 13.51  \\ 
 \hline
 Lima  & 6.28 / 6.73 / 7.15 & 6.36 / 12.62 / 19.21 \\
 \hline
 Quito & 4.21 / 4.56 / 4.75 & 4.47 / 8.42 / 13.04  \\
 \hline
 Jakarta & 5.85 / 6.14 / 6.16  & 6.01 / 11.72 / 16.98 \\
 \hline
\end{tabular}
 \caption{Experiments on IBM-Q.}
 \label{table2}
\end{table}
\section{Discussion and Outlook}

In this project, we study a quantum index search problem within a quantum-classical system. Based on Grover's algorithm, we propose \sol~ that queries quantum computer iteratively and process the quantum results on the classical part. With the assistance of classical computers, \sol~ can reduce the input set for each query. Consequently, \sol~ requires fewer qubits. With this iterative qubit management, \sol~ reduces qubit consumption, up to 66.2\%, with a reasonable accuracy compared with GSearch. 
Our work provides a general step forward in quantum resource management for the future hybrid quantum cloud era. With the improved consumption, the limited qubits are possible to be shared with other users in a multi-tenant architecture. 
%%%%%%%%%%%%%%%%%%%%%%

There is, however, still significant progress to be made in this domain. \sol~ algorithms work on classical computers. In \sol, the threshold-based Algorithm~\ref{alg:search} may suffer from true negative scenarios, where targets are filtered out without a recovery mechanism. A potential improvement could be adding redundancies from $NS$ set at each iteration. Additionally, \sol~, the heuristic system, lacks of theoretical analysis and proved performance boundaries. Experiments on IBM-Q show that 
the initialization of each quantum query is an expensive operation in the current NISQ era. 
Efficient collaboration and task distribution between quantum and classical computers in a hybrid cluster should be intensively investigated.

%Depending on the threshold value, \sol~ may suffer from true negative scenarios, where targets are filtered out without a recovery mechanism. 

% \fi
%\IEEEpeerreviewmaketitle

%\newpage

\bibliography{routing}
\bibliographystyle{IEEEtran}

\end{document}